\documentclass[final,5p,times,twocolumn]{elsarticle}

\usepackage{lineno}
\usepackage{graphicx}
\usepackage{amssymb}
\usepackage{pdflscape}
\usepackage[english]{babel}
\usepackage[dvipsnames,svgnames,x11names]{xcolor}
\usepackage{booktabs,tabularx}
\usepackage{multirow}
\usepackage{subfig} 
\usepackage{hyperref}
\usepackage{amsmath}
\usepackage{commath}
\usepackage[english]{babel}
\usepackage[ruled]{algorithm2e}
\SetEndCharOfAlgoLine{}
\usepackage[leftmargin=6em,rightmargin=12em,indentfirst=false]{quoting}
\usepackage{siunitx}

\usepackage[final]{changes}
\usepackage{float}

\usepackage{lineno}
\usepackage{tikz}
\usepackage[export]{adjustbox}

\usepackage{graphicx}
\usepackage[utf8]{inputenc}
\usepackage[export]{adjustbox}
\usepackage{wrapfig}
\usepackage{dcolumn}

\makeatletter
\newcolumntype{T}[3]{>{\textfont0=\the@{#1}{#2}{#3}}c<{\DC@end}}
\makeatother

\usepackage{pgfplots}
\pgfplotsset{width=10cm,compat=1.9}

\usepackage{array}

\newcolumntype{L}[1]{>{\raggedright\let\newline\\\arraybackslash\hspace{0pt}}m{#1}}
\newcolumntype{C}[1]{>{\centering\let\newline\\\arraybackslash\hspace{0pt}}m{#1}}
\newcolumntype{R}[1]{>{\raggedleft\let\newline\\\arraybackslash\hspace{0pt}}m{#1}}

\usepackage{subfiles}

\usepackage{todonotes}

\journal{Indoor Air 2022 Conference}
\begin{document}
	
\begin{frontmatter}

\title{Smartwatch-based ecological momentary assessments for occupant wellness and privacy in buildings}

\author{Clayton Miller\,$^{1,*}$, Renee Christensen$^{2}$, Jin Kai Leong$^{1}$, Mahmoud Abdelrahman$^{1}$, Federico Tartarini$^{3}$, Matias Quintana$^{1}$, Andre Matthias Müller$^{4}$, Mario Frei$^{1}$}

\address{$^{1}$College of Design and Engineering, National University of Singapore (NUS), Singapore}
\address{$^{2}$International Well Building Institute (IWBI), New York, NY, USA}
\address{$^{3}$Berkeley Education Alliance for Research in Singapore (BEARS), Singapore}
\address{$^{4}$Saw Swee Hock School of Public Health, National University of Singapore (NUS), National University Health System, Singapore}
\address{$^*$Corresponding Author: clayton@nus.edu.sg, +65 81602452}

\begin{abstract}
This paper describes the adaptation of an open-source ecological momentary assessment smart-watch platform with three sets of micro-survey wellness-related questions focused on i) infectious disease (COVID-19) risk perception, ii) privacy and distraction in an office context, and iii) triggers of various movement-related behaviors in buildings. This platform was previously used to collect data for thermal comfort, and this work extends its use to other domains. Several research participants took part in a proof-of-concept experiment by wearing a smartwatch to collect their micro-survey question preferences and perception responses for two of the question sets. Participants were also asked to install an indoor localization app on their phone to detect where precisely in the building they completed the survey. The experiment identified occupant information such as the tendencies for the research participants to prefer privacy in certain spaces and the difference between infectious disease risk perception in naturally versus mechanically ventilated spaces.
\end{abstract}


\end{frontmatter}


\section{Introduction}
\label{sec:introduction}

When rethinking urban planning and building designs, practitioners need to evaluate what features of the current built environment context contribute to the health and well-being of its occupants~\cite{OBrien2020-ns,Altomonte2020-rh,Stazi2017-ep}. One example of this challenge is the mitigation of airborne infectious diseases such as COVID-19. Designers or operators would want to understand, for example, which building features could be intersection zones that cause congestion or to determine which high-touch building features occupants perceive as a risk of transmission~\cite{Kim2021-gh}). Another often overlooked aspect is understanding what building features relate to the feeling of privacy and distraction. It has been established that the conventional open plan office space provides many opportunities for people to feel a lack of privacy in addition to visual and noise-based distractions from those around them~\cite{Kim2013-gs}. Finally, with the emerging emphasis on wellness in the built environment, buildings are designed with indirect encouragements to nudge occupants to make decisions that will improve their health~\cite{Timm2018-wf}. An example is when staircases are made more prominent and visually appealing to encourage occupants to use them rather than elevators or lifts~\cite{Potrc_Obrecht2019-jd}. Post-occupancy evaluations can help collect data from occupants to understand better how they perceive their environment; however, they are limited in the amount and frequency of data they can collect. Wearable technologies can be used to fill this gap by capturing feedback at high spatial and temporal resolutions that can improve these design-related efforts~\cite{Li2018-yz}.

The Cozie platform was developed to facilitate repeated right-here-right-now surveys across extended periods of time using micro-ecological momentary assessments~\cite{Jayathissa2019-kg}. The platform proved to be an efficient way to collect subjective feedback as compared to other post-occupancy feedback methods such as smartphone-based surveys. Subsequent field studies evaluated the platform’s ability to capture a large number of thermal, aural, and visual comfort feedback to create personalized comfort models~\cite{Jayathissa2020-pv}, the ability to create personal comfort models using spatial proximity~\cite{Abdelrahman2022-zs} and understand the thermal comfort transition behavior of occupants between different spaces~\cite{Sae-Zhang2020-fy}. Cozie is open source and can be easily modified to include different sets of questions beyond thermal comfort. Two versions of Cozie have been developed for both the Fitbit and Apple Watch platforms. This paper outlines three survey flows for Cozie Fitbit developed to capture occupant preference information related to the perceived risk of infectious diseases, privacy and distraction in an office context, and the potential triggers of various movement-related behaviors in the built environment.

\begin{figure*}[!h]
	\centering
	\includegraphics[width=0.9\linewidth]{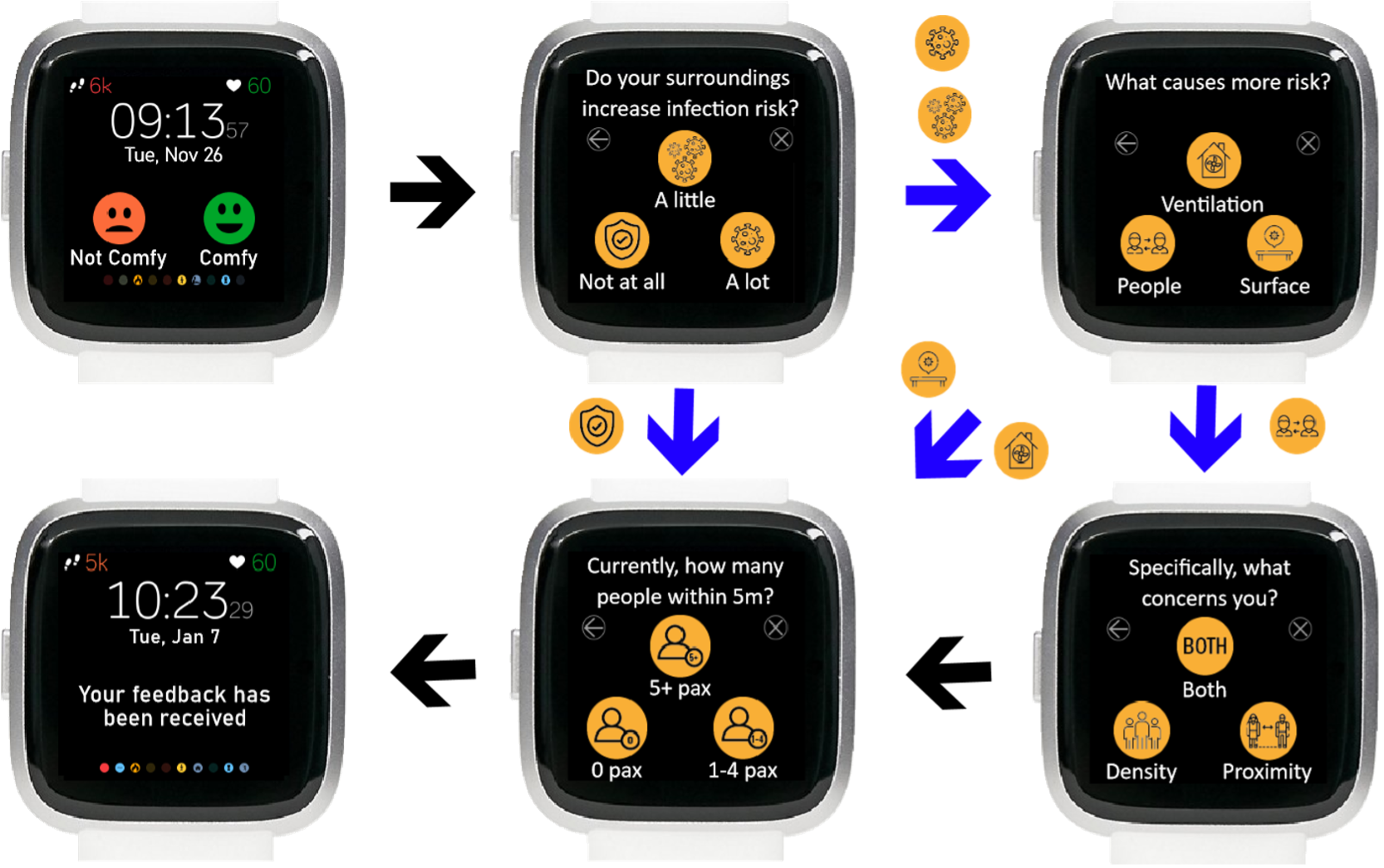}
	\caption{Infectious disease risk perception micro-survey flow overview. The arrows indicate the flow of questions with a black arrow indicating a single path and a blue arrow indicating multiple paths based on the option selected.}
	\label{fig:infectionrisk}
\end{figure*}

\begin{figure*}[!h]
	\centering
	\includegraphics[width=0.75\linewidth]{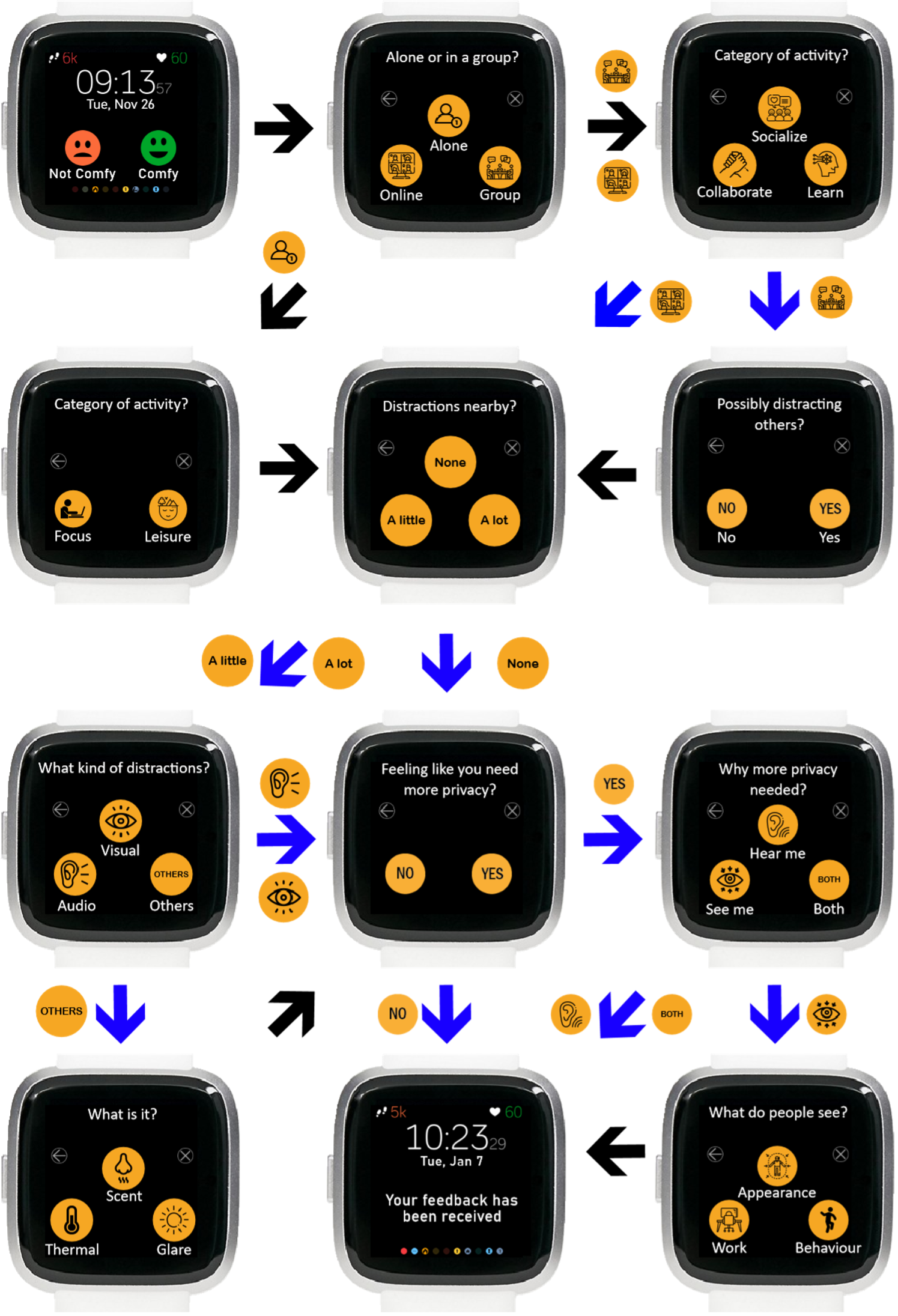}
	\caption{Privacy, distraction, and surroundings impact micro-survey question flow overview.}
	\label{fig:privacy}
\end{figure*}

\begin{figure*}[!h]
	\centering
	\includegraphics[width=0.95\linewidth]{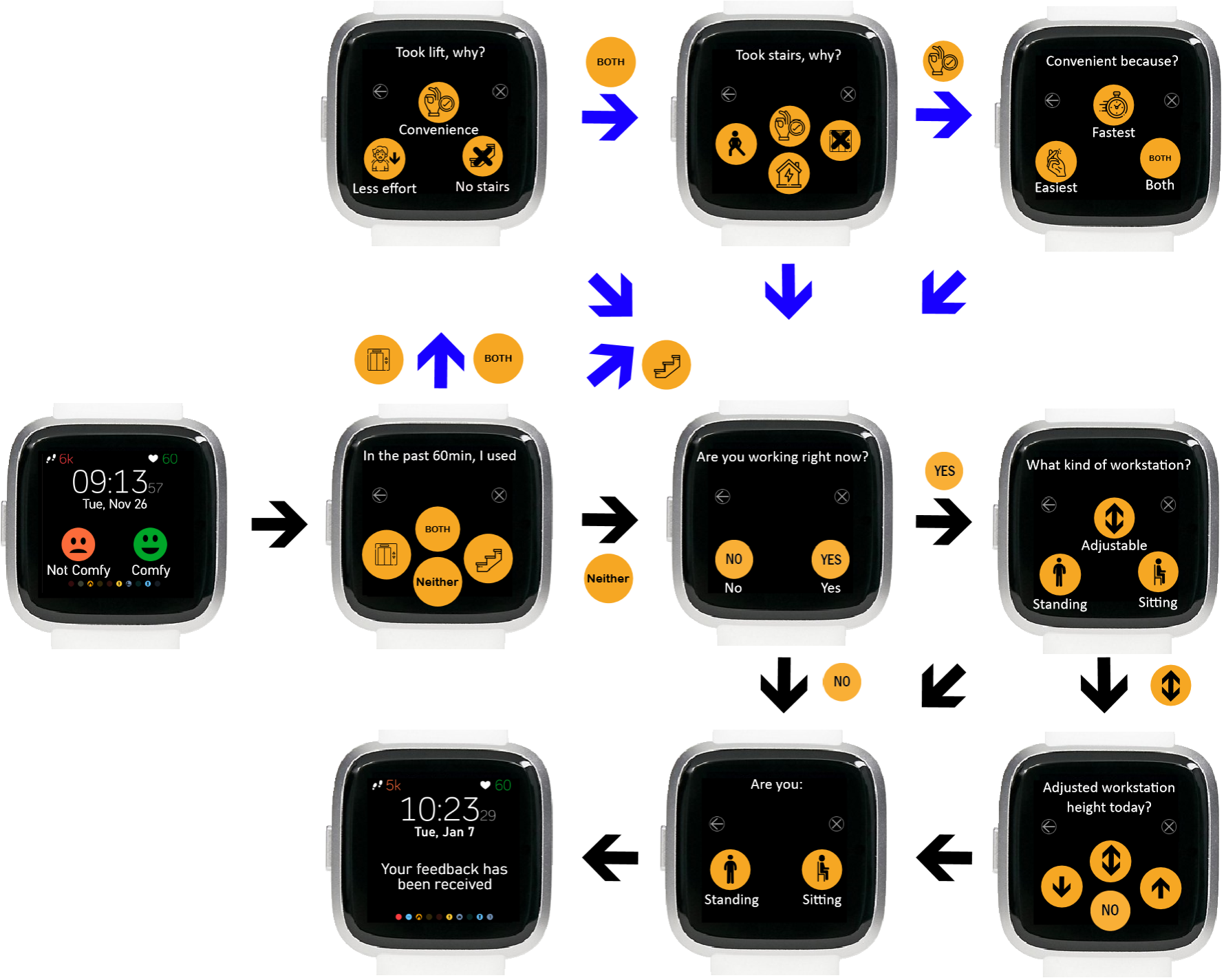}
	\caption{Evaluation of movement-related decisions micro-survey question flow overview.}
	\label{fig:movement}
\end{figure*}

\section{Methods}
This paper outlines the deployment of three micro-survey question sets, of which, two are used in a pilot longitudinal study. These sets include 2-7 multiple-choice questions that are answered using the Cozie watch-face installed on a Fitbit smartwatch. The surveys have a built-in logic that determines which follow-up question is presented to a user based on their previous answers. The following subsections outline the three question flows tested in this study. These surveys were implemented in the open-source Cozie Fitbit project hosted on a public repository on GitHub\footnote{\url{https://github.com/cozie-app/cozie}}.

\subsection{Infectious disease risk perception}
The first question set relates to an occupant’s perception that their surroundings impact their risk of acquiring an infectious disease, as shown in Figure \ref{fig:infectionrisk}. This perception is essential in evaluating whether there are perceived risks to occupants that could be mitigated by changing some building features. It also identifies occupant risk perception that may not align with the science of disease spread. These off-target concerns could then be assuaged through occupant education. The user first starts from the home screen on the upper left and answers the core question related to their perception of infection risk at that moment and location. Two follow-up questions evaluate what aspect of risk is specifically concerning for them. Finally, there is a question to request the number of people within a 5-m radius.

\subsection{Privacy and distraction perception}
The next question set, seen in Figure \ref{fig:privacy}, focuses on capturing whether occupants feel like they need more privacy and whether they find their surroundings distracting. The survey begins with a question about whether someone is in a group or alone and what category of activity they are undertaking at that point in time. These foundational questions establish the context for the subsequent ones about distractions. A person working alone has different focus, sound, and visual privacy needs than a group of people. The following questions focus on whether the person feels as if they need more privacy, and if so, what are the specific concerns and what are they worried about in terms of others around them. These follow-up questions dive into whether the person needs privacy because they worry about others seeing and/or hearing their appearance, work, or behavior.

\subsection{Triggers related to movement in buildings}
The last set of questions, shown in Figure \ref{fig:movement}, captures the effectiveness of nudging occupants to be more active in buildings, i.e., the use of stairways instead of elevators/lifts and adjustable height workstations. These activities are design-driven, and evaluation of users’ actual behavior in buildings with these features informs about the adoption and need for education, signage, and placement. This flow begins with a question asking whether the occupant has used the stairs or lift in the building in the last one hour and, based on the response, seeks to determine the reason for that decision. The flow then moves to a question about if and how the occupant is using an adjustable-height workstation. This question set is included for schematic purposes, but data from this flow are not included in this paper.

\subsection{Deployment in a university building context}
The proof-of-concept test deployment for this paper included six participants who were asked to wear a Fitbit smartwatch and use the Cozie application. Only data for the first two question sets are included in this study. The watch generated a vibration-based prompt for feedback every 1, 2, or 3 hours between 9 am and 9 pm. Each question set was expected to take between 3-9 seconds to complete, and only one set was deployed at a time. Participants were able to provide feedback outside of the prompts, provided that two consecutive responses were more than 15 minutes apart. They were required to install an indoor localization app that uses Bluetooth signals from beacons to track their movements within the case study buildings. The methodology for the first two question sets was approved by the NUS Institutional Review Board (IRB) (NUS-IRB-2020-135), and participants were incentivized with an SG\$50 voucher.

\section{Results and discussion}
\label{sec:results}

\begin{figure}[!h]
	\centering
	\includegraphics[width=\linewidth]{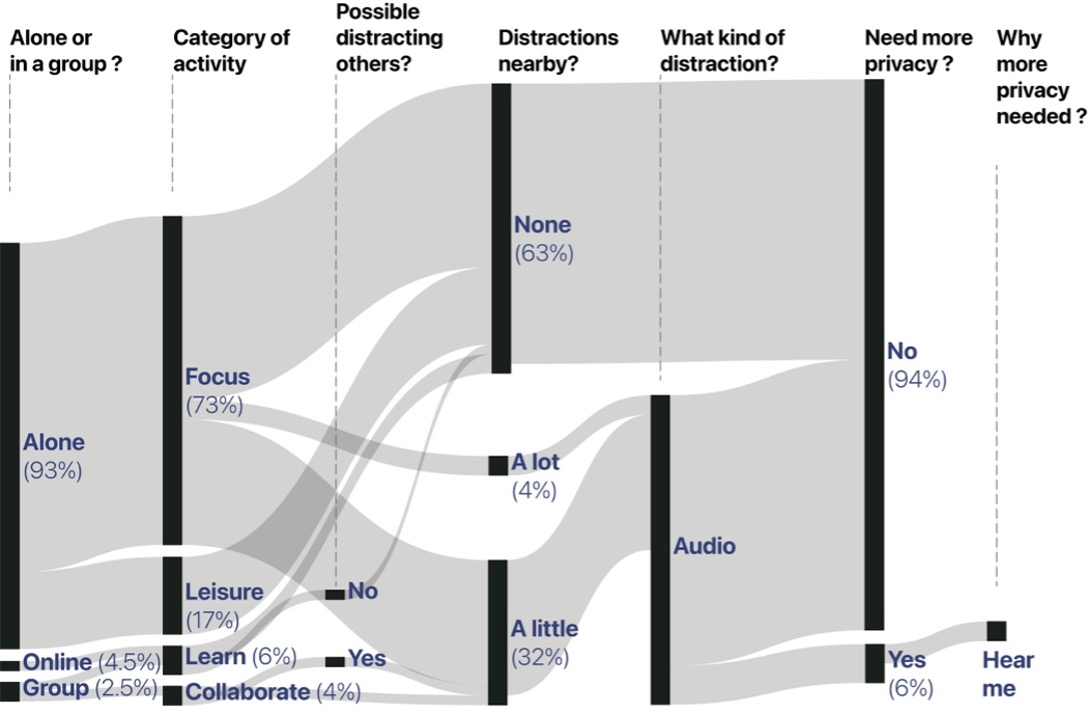}
	\caption{Evaluation of movement-related decisions micro-survey question flow overview.}
	\label{fig:privacydata}
\end{figure}

\begin{figure}[!h]
	\centering
	\includegraphics[width=\linewidth]{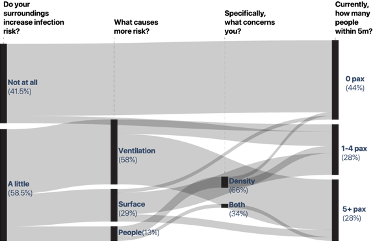}
	\caption{Evaluation of movement-related decisions micro-survey question flow overview.}
	\label{fig:infectiondata}
\end{figure}

The data collection for this experiment resulted in subjective feedback from six participants over the span of 30 days. Figure \ref{fig:privacydata} illustrates aggregations of this feedback from the Privacy and Distraction question flows. For this data set, most of the activity of the test subjects was solitary, with a strong emphasis on the occupant focusing. Less than half the time, there were distractions nearby, with the most significant reason being noise. Despite the distraction, the subjects felt as if they needed more privacy only 6\% of the time, and the primary source of that feeling was requiring more protection from people seeing their work. Figure \ref{fig:infectiondata} shows the aggregated results and Infectious Disease Risk Perception question flow. For this data set, participants felt for a slim majority of the time that there was an increase of risk of infection, with ventilation being the most prominent concern followed by surfaces and then people density.

\section{Conclusion}
\label{sec:conclusion}
This paper outlines the development of micro-ecological momentary assessments on a smartwatch to characterize occupant preference in wellness and privacy topics that have been previously unexplored. An experiment was conducted that collected spatially and temporally diverse preference data from real-world occupants of a case study building. The results gave insight into whether privacy and distraction were issues and how much perceived risk of infectious diseases was present. In future work, larger case study deployments across different buildings with a diversity of design objectives could be used to compare design features in the context of these three wellness and satisfaction objectives. In addition, there are opportunities to further analyze which spatial regions are better or worse for the objective of each question flow and why. The anonymized data set and example code for this paper can be found in a GitHub repository\footnote{\url{https://github.com/buds-lab/ema-for-occupant-wellness-and-privacy}}. The Cozie platform is available for both Fitbit smartwatches\footnote{\url{https://cozie.app/}} and Apple Watches\footnote{\url{https://www.cozie-apple.app/}} and is designed to be an open-source, community-driven effort for others to collaborate.

\section*{Acknowledgements}
The author team would like to thank several student researchers who contributed to the data collection process for this publication including Charis Boey Shand Yin, Chua Yun Xuan, Ang Si Hui Pearlyn, and Tan Jing En Charlene. This research was funded by the Singapore Ministry of Education (MOE) through the Tier 1 Grants: Ecological Momentary Assessment (EMA) for Built Environment Research (A-0008301-00-00) and The Internet-of-Buildings (IoB) Platform – Visual Analytics for AI Technologies towards a Well and Green Built Environment (A-0008305-00-00). The Republic of Singapore’s National Research Foundation (NRF) through the SinBerBEST program provided partial support for manpower for this project.

\bibliographystyle{model1-num-names}
\bibliography{references}

\end{document}